\def\red#1{#1}

\newcommand{\beq}{\begin{equation}}
\newcommand{\eeq}{\end{equation}}
\newcommand{\beqa}{\begin{eqnarray}}
\newcommand{\eeqa}{\end{eqnarray}}

\newcommand{\opx}{\hat{\sigma}_x}
\newcommand{\opy}{\hat{\sigma}_y}
\newcommand{\opz}{\hat{\sigma}_z}
\newcommand{\identity}{\hat{\openone}}

\newcommand{\ot}{\otimes}
\newcommand{\opcz}{\hat{Z}_c}
\newcommand{\opup}{\hat{U}_p}
\newcommand{\opuc}{\hat{U}_c}
\newcommand{\opur}{\hat{U}_r}

\newcommand{\opxii}{\hat{\sigma}_x}
\newcommand{\opyii}{\hat{\sigma}_y}
\newcommand{\opzii}{\hat{\sigma}_z}
\newcommand{\identityii}{\hat{\openone}}
\newcommand{\opxiii}{\hat{\sigma}_x}
\newcommand{\opyiii}{\hat{\sigma}_y}
\newcommand{\opziii}{\hat{\sigma}_z}
\newcommand{\identityiii}{\hat{\openone}}
\newcommand{\opxiv}{\hat{\sigma}_x}
\newcommand{\opyiv}{\hat{\sigma}_y}
\newcommand{\opziv}{\hat{\sigma}_z}
\newcommand{\identityiv}{\hat{\openone}}

\documentclass[pra,superscriptaddress,showpacs,preprint,eqsecnum]{revtex4}

\usepackage{graphicx}

\begin{document}

\title{On the structure of the sets of \red{mutually} unbiased bases
for \red{$N$ qubits}}

\author{\red{J. L. Romero}}
\affiliation{School of \red{Information and Communication}
Technology, Royal Institute of Technology (KTH), Electrum 229,
SE-164 40 Kista, Sweden}

\author{G. Bj\"{o}rk}
\affiliation{School of \red{Information and Communication}
Technology, Royal Institute of Technology (KTH), Electrum 229,
SE-164 40 Kista, Sweden}

\author{A. B. Klimov}
\affiliation{Departamento de F\'{\i}sica, Universidad de
Guadalajara, 44420~Guadalajara, Jalisco, Mexico}

\author{L. L. S\'anchez-Soto}
\affiliation{Departamento de \'{O}ptica,
Facultad de F\'{\i}sica, Universidad Complutense,
28040~Madrid, Spain}

\date{\today}

\begin{abstract}
\red{For a system of $N$ qubits}, spanning a Hilbert
space of dimension \red{$d=2^N$}, it is known that
there exists $d+1$ \red{mutually unbiased bases}.
Different construction algorithms exist, and it is
remarkable that \red{different} methods lead to
sets of bases with different properties \red{as
far as separability is concerned}. Here we derive
\red{the} four sets of nine bases for three qubits,
and show how they are unitarily related. We also
briefly discuss the \red{four-qubit} case, give
the entanglement structure of \red{sixteen} sets
of bases,and show some of them, and their
interrelations, as examples. \red{The extension
of the method to the general case of $N$ qubits
is outlined.}
\end{abstract}

\pacs{3.65.Ta, 03.65.Ud, 03.65.Ca}

\maketitle

\section{Introduction}

Every quantum system is associated with some state
\red{(pure or mixed)} in a Hilbert space. \red{It is
possible to ascertain this quantum state by performing
a series of measurements on an ensemble} consisting
of many identically members. Each measurement will
\red{modify} the measured ensemble member in such
way that it\red{, in general,} not possible to get any
additional information about the original \red{state}.
Several techniques, such as state tomography~\cite{Vogel,Leonhardt},
maximum likelihood~\cite{Helstrom,Banaszek,Hradil}\red{,}
and maximum\red{-}entropy methods~\cite{Slater} (\red{or}
combinations thereof~\cite{Rehacek}), have been
\red{devised} for state estimation.

When the Hilbert space is finite\red{,} it has been shown
that the optimal approach to get the information is
related to \red{a} special set of states \red{that} are
\red{``mutually unbiased"}~\cite{Ivanovic,Wootters,Wootters 2},
\red{for which the uncertainty spread of the inferred
state is minimized}. \red{Note, however, that} we are
ignoring more general measurements\red{,} such as joint
measurements \red{on} all \red{the} members of the
ensemble~\cite{Peres,Massar,Derka} \red{or} adaptive
measurements~\cite{Brody,Fisher}, \red{which} surpass
the ability of {\em a priori} fixed, single\red{-}copy
measurements.

\red{Let us denote basis sets by $A = 1,2, \ldots$ and states
within a basis by $|A, a \rangle$, with $a = 1, 2, \ldots,d,$ $d$
being the dimension of the Hilbert space. We recall that two bases
$| A, a \rangle$ and $|B, b\rangle$ are said to be mutually
unbiased bases (MUBs) if a system prepared in any element of $A$
has a uniform probability distribution of being found in any
element of B, that is} \beq \red{ | \langle A, a | B, b \rangle|^2
= 1/d ,} \eeq where orthonormality among states of the same basis
is assumed. These \red{MUBs} are central to the formulation of the
\red{discrete} Wigner function~\cite{Wootters 0,Asplund
2,Gibbons,Vourdas}. They have also have also been used in
\red{cryptographic} protocols~\cite{Cryptography,Asplund}\red{,}
due to the complete uncertainty about the outcome of a measurement
in some basis after the preparation of the system in another, if
the bases are mutually unbiased. MUBs are also used for quantum
error correction codes~\cite{Gottesman,Calderbank} \red{and
recently} they have also found uses in quantum game theory, in
particular to provide a solution to the \red{``mean king
problem''} \cite{Aharonov,Englert,Aravind,Hayashi,Durt,Paz}.

It has been shown that the \red{maximum number} of MUBs can be at
most $d+1$ \cite{Ivanovic}. Actually, it is known that if $d$ is
prime or \red{power of prime,} the maximal number of MUBs can be
achieved~\cite{Ivanovic,Wootters 2}. \red{Remarkably though, there
is no known answer for any other values of $d$,} although there
are some approaches to try to find the solution to this problem in
some simple cases, such as $d=6$ or when $d$ is an nonprime
integer squared~\cite{Grassl,Archer,Wocjan}. \red{Recent works
have suggested that the answer to this question may well be
related with the non-existence of finite projective planes of
certain orders~\cite{Saniga04,Bengtsson04} or with the problem of
mutually orthogonal Latin squares in
combinatorics~\cite{Zauner99,Wootters04b}.}

Experimental quantum information and computation have already
moved from \red{single-qubit} protocols to several qubits (at
present around eight~\cite{Blatt}), so there is a need to extend
our knowledge, \red{especially} about entanglement properties of
several qubits. This also includes extensions of measurement
techniques of systems with more than \red{two qubits}. Therefore,
a new problem related with \red{MUBs naturally} appears, namely
that for more than two qubits, different MUB structures exist,
where the word ``structure'' refers to the entanglement properties
of the bases. We are already aware of the existence of three MUB
structures for three qubits~\cite{Wootters 2,Lawrence}. In this
paper we will show that, in fact, there exists exactly four
different MUB structures in this space. We will also show how they
are interrelated. For the experimentalist, this information is
\red{very} important, because the complexity of an implementation
of two or more \red{MUBs} will\red{,} of course\red{,} greatly
depend on how many of the qubits that need to be entangled. We
will also briefly discuss the \red{four-qubit} case and show that
in this space there exist \red{sixteen} different MUB structures.
We will \red{exhibit} which they are and derive some of them
explicitly. It is then possible to continue and analyze the
general \red{$N$-}qubit MUBs much in the same manner,
although\red{, for brevity and simplicity,} we will stop at four
qubits.

\section{MUBs for one and two qubits}

Because states belonging to \red{the} same basis are usually taken
to be orthonormal, to study the property of "mutually
unbiasedness" it is possible to use either mutually unbiased bases
or the operators which have the basis states as eigenvectors. We
\red{thus} need $d^2-1$ operators to obtain the whole set of
states. In the case of prime and power of a prime dimensions this
set of operators can be collected as $d+1$ sets of $d-1$ commuting
operators, which is related with the grading of a Lie
algebra~\cite{Patera}. In the case of $N$-qubits
($2^N$-dimensional case) we need $N$ commuting operators to define
uniquely a pure state~\cite{Zeilinger,Brukner}.

In finite dimensional systems is also possible to define a
discrete phase space, and when the dimension of the system is
either a prime or a power of prime the phase space is a finite
geometry. The above operators are related with translations in
this phase space and (without a phase factor) they are the
so\red{-}called displacement operators\red{,} which satisfy the
covariant property of the discrete Wigner function defined
there~\cite{Gibbons}.

The \red{two}-dimensional Hilbert vector space
(\red{one} qubit) is spanned, e.g., by the two
orthonormal eigenvectors \red{of} the spin 1/2
observable $\opz$\red{,} which will be used,
in the following, as our computational basis.

In this \red{Hilbert} space, the MUB set of $2^2-1 = 3$ bases is
given by the eigenvectors of the Pauli matrices $\opx$,
$\opy$\red{,} and $\opz$. Any unitary operation \red{preserves}
the angles between the axes of the transformed operators, so we
can redefine our coordinates to have a new set of Pauli matrices.
\red{We can then say} that the structure of the MUBs remains
invariant under any unitary transformation. This is akin to saying
that only one MUB structure exists in the two-dimensional Hilbert
space.

The same result also holds for two qubits, although, in addition,
the extra feature of entanglement appears. Several \red{methods}
have been presented for the \red{explicit} construction of
MUBs~\cite{Ivanovic,Wootters
2,Wocjan,Calderbank,Chaturvedi,Pittenger,Durt}. Here we will only
focus on one of them, \red{which is based on the use of the finite
Fourier transform, employing} the operators $\opx$, and $\opz$ and
tensor products~\cite{Bandyopadhyay,Klimov}. (Because this work
attempts to delineate the structure and interrelation between MUBs,
and not their explicit mathematical construction, we omit such a
discussion and direct interested readers to the aforementioned
work.) If we follow \red{the algorithm} in \cite{Klimov}, we get a
table with five rows of three mutually commuting (tensor products
of) operators, shown in Table~\ref{table0}. (We have suppressed the
tensor multiplication sign in all the tables.) The \red{table}
reproduces Eqns.~(3.30) and (3.32)-(3.335) in
\red{Ref.~\cite{Klimov}}.

\red{By construction, the algorithm guarantees that}
the simultaneous eigenstates of the operators in each
row give a complete basis, \red{and} each basis is
mutually unbiased to each other. \red{The} number
on the left enumerates the bases, \red{while}
the number on the right denotes how many subsystems
the bases can be factorized into.

It is easy to see that the three first bases are fully separable
(the three operators in each of the first three rows commute for
each of the two subsystems, separately), and that the last two
bases are not separable. In fact, their simultaneous eigenstates
are all maximally entangled states. We call this MUB construction
with three (bi-)separable and two nonseparable MUBs a (3,2)
construction. This \red{is equivalent,} under local unitary
transformations\red{, to} the construction given in Fig.~1 \red{of
Ref.~\cite{Lawrence}.}

\begin{table}[h]
\red{
\begin{tabular}{||c||c|c|c||c||} \hline
1 & $\opz \opz $ & $\identity \opz$ &
$\opz \identity$ & 2\\
\hline 2 & $\opx \opx $ & $\identity  \opx $ &
$\opx \identity$ & 2\\
\hline
3 & $\opy \opy$ & $ \identity \opy $ &
$\opy  \identity $ & 2\\
\hline
4 & $\opx \opy$ & $\opz \opx $ &
$\opy \opz $ & 1\\
\hline
5 & $\opy \opx $ & $\opz  \opy $ &
$ \opx \opz$  & 1 \\
\hline
\end{tabular}}
\caption{Five sets of three operators defining a (3,2) MUB.}
\label{table0}
\end{table}

The \red{algorithm} imposes several characteristic
features of the operator table. The table is composed of
binary tensor products of the four operators $\opz$, $\opx$,
$\opy=i \opx \opz = -i \opz \opx$, and $\identity = \opx^2 =
\opy^2=\opz^2$. In all, there exists $4^2=16$ combinations
of such products, but the operator $\identity \ot \identity $
must be excluded because \red{it} commutes with every
operator in the set. Each of the $15=3 \cdot 5$
remaining operators is represented once in the table
above. Moreover, the table is uniquely defined by the
four entries in the two first columns of the first two
rows. All other operators $O_{r,c}$ are \red{determined}
by the relations $O_{r,c}=O_{r,c-2}O_{r,c-1}$,
and $O_{r,c}= O_{2,c} O_{1,c+r-3}$ for $r>2$,
where \red{the indices $r$ and $c$} denote the
row and the column of the operator, respectively,
and \red{must} be taken \red{modulo} four.

Noting that each separable basis (i.e., the first
three rows) has two eigenoperators containing the
\red{identity}, that a nonseparable basis cannot
have any eigenoperator containing the identity,
and that there must be six entries, $\identity \ot \opx,
\identity \ot \opy, \ldots , \opy \ot \identity, \opz \ot
\identity$, containing the identity in the table, we can
conclude that the (3,2) \red{is} the only possible
construction in this space. That is, any unitary
transformation, local or nonlocal, will yield an
isomorphic table with respect to the separability,
\red{except, perhaps, for some row permutations.}

\section{MUB structures for three qubits}

\red{Lawrence, Brukner, and Zeilinger~\cite{Lawrence}} have
\red{shown} explicitly two different sets of \red{MUBs} in the
case $d=8$. One of them has three fully separable bases (every
eigenvector of these three bases is a tensor product of states
embedded in the Hilbert space of each \red{single} qubit) and six
GHZ bases~\cite{Greenberger}. \red{The} other structure has nine
sets of bases with eigenvectors where \red{one qubit} can be
factorized and the other two qubits are in a maximally entangled
state. If we follow \red{again the algorithm in
Ref.~\cite{Klimov}} we get Table~\ref{table1}.

\begin{table}[h]
\begin{tabular}{||c||c|c|c|c|c|c|c||c||} \hline
1 & $\opz \identity \identity$ & $\identity \identity \opz$ &
$\identity \opz \identity$ & $\opz \identity \opz$ & $\identity
\opz \opz$ & $\opz \opz \opz$ & $\opz \opz \identity$ & 3 \\
\hline
2 & $\opx \identity \identity$ & $\identity \opx \identity $ &
$\identity \identity \opx$ & $\opx \opx \identity$ & $\identity
\opx \opx$ & $\opx \opx \opx$ & $\opx \identity \opx$ & 3 \\
\hline
3 & $\opy \identity \identity$ & $\identity \opx \opz$ &
$\identity \opz \opx$ & $\opy \opx \opz$ & $\identity \opy \opy$ &
$\opy \opy \opy$ & $\opy \opz \opx$ & 2 \\
\hline
4 & $\opx \identity \opz$ & $\identity \opy \identity$ &
$\opz \identity \opy $ & $\opx \opy \opz$ & $\opz \opy \opy$ &
$\opy \opy \opx$ & $\opy \identity \opx$ & 2 \\
\hline
5 & $\opx \opz \identity$ & $\opz \opx \opz$ &
$\identity \opz \opy$ & $\opy \opy \opz$ &
$\opz \opy \opx$ & $\opy \opx \opx$ & $\opx \identity \opy$ & 1\\
\hline
6 & $\opy \identity \opz$ & $\identity \opy \opz$ &
$\opz \opz \opy$ & $\opy \opy \identity$ &
$\opz \opx \opx$ & $\opx \opx \opy$ &
$\opx \opz \opx$ & 1 \\
\hline
7 & $\opx \opz \opz$ & $\opz \opy \opz$ &
$\opz \opz \opx$ & $\opy \opx \identity$ &
$\identity \opx \opy$ & $\opx \opy \opx$ &
$\opy \identity \opy$ & 1\\
\hline
8 & $\opy \opz \opz$ & $\opz \opy \identity$ &
$\opz \identity \opx$ & $\opx \opx \opz$ &
$\identity \opy \opx$ & $\opy \opx \opy$ &
$\opx \opz \opy$ & 1 \\
\hline
9 & $\opy \opz \identity$ & $\opz \opx \identity $ &
$\identity \identity \opy$ & $\opx \opy \identity$ &
$\opz \opx \opy$ & $\opx \opy \opy$ & $\opy \opz \opy$ & 2 \\
\hline
\end{tabular}
\caption{Nine sets of operators defining a (2,3,4) MUB.}
\label{table1}
\end{table}

Table~\ref{table1} is equivalent to \red{the first MUB
construction demonstrated in this space} by Fields and
Wootters~\cite{Wootters 2} or \red{to the example~2} of Sec.~5 in
Ref.~\cite{Pittenger}, in that it has two fully (that is, tri-)
separable bases (marked with a 3 in the rightmost column), three
biseparable bases (marked with a 2), and four nonseparable bases
(marked with a 1). We will denote such a set of MUBs \red{as a}
(2,3,4) \red{structure}.

Lawrence, Brukner, and Zeilinger~\cite{Lawrence} \red{have}
pointed out that two other constructions are possible, namely a
\red{(3,0,6) set} of MUBs where three of the bases are fully
separable, and the remaining six bases are nonseparable, and
\red{a (0,9,0) in which all} the bases are biseparable. The
\red{corresponding operators} are given in Figs.~2 and 4 \red{of}
Ref.~\cite{Lawrence}. We would like to see how we can \red{derive}
these bases from \red{the ones in} Table~\ref{table1}. To this end
we use the controlled-$Z$ operator \beq \opcz = \left (
\begin{array}{cccc}
1 & 0 & 0 & 0 \\
0 & 1 & 0 & 0 \\
0 & 0 & 1 & 0 \\
0 & 0 & 0 & -1
\end{array}
\right ) . \eeq This operator is unitary, nonseparable and,
moreover, has the property that it is its own inverse,
$\opcz^{-1}=\opcz$, and its own conjugate. It commutes with the
operators $\identity \otimes \opz$, $\opz \otimes \identity$, and
$\opz \otimes \opz$. Let us first convert the (2,3,4) \red{into}
an equivalent basis set by applying the local unitary
(permutation) transformation $x \rightarrow y \rightarrow z
\rightarrow x$ to the leftmost qubit. The operator performing this
transformation (up to an overall phase-factor) is \beq \opup =
\frac{1}{\sqrt{2}} \left (
\begin{array}{cc}
e^{i \pi/4} & e^{i 3 \pi/4}  \\
e^{i \pi/4} & e^{-i \pi/4}
\end{array}
\right ) .
\eeq
We also apply the permutation $y \leftrightarrow z$ to
the middle and rightmost qubits. The corresponding
\red{operator} is
\beq
\opuc =
\frac{1}{\sqrt{2}}
\left (
\begin{array}{cc}
1 & i  \\
i & 1
\end{array}
\right ) .
\eeq
Applying the operator $\opup \otimes \opuc \otimes \opuc$
to Table~\ref{table1} above, we are left with an
equivalent operator table, Table~\ref{table2}, still
defining a (2,3,4) MUB.

\begin{table}[h]
\begin{tabular}{||c||c|c|c|c|c|c|c||c||} \hline
1 & $\opx \identity \identity$ & $\identity \identity \opy$ &
$\identity \opy \identity$ & $\opx \identity \opy$ &
$\identity \opy \opy$ & $\opx \opy \opy$ &
$\opx \opy \identity$ & 3 \\
\hline
2 & $\opy \identity \identity$ & $\identity \opx \identity $ &
$\identity \identity \opx$ & $\opy \opx \identity$ &
$\identity \opx \opx$ & $\opy \opx \opx$ &
$\opy \identity \opx$ & 3 \\
\hline
3 & $\opz \identity \identity$ & $\identity \opx \opy$ &
$\identity \opy \opx$ & $\opz \opx \opy$ &
$\identity \opz \opz$ & $\opz \opz \opz$ &
$\opz \opy \opx$ & 2 \\
\hline
4 & $\opy \identity \opy$ & $\identity \opz \identity$ &
$\opx \identity \opz $ & $\opy \opz \opy$ &
$\opx \opz \opz$ & $\opz \opz \opx$ &
$\opz \identity \opx$ & 2 \\
\hline
5 & $\opy \opy \identity$ & $\opx \opx \opy$ &
$\identity \opy \opz$ & $\opz \opz \opy$ &
$\opx \opz \opx$ & $\opz \opx \opx$ &
$\opy \identity \opz$ & 1 \\
\hline
6 & $\opz \identity \opy$ & $\identity \opz \opy$ &
$\opx \opy \opz$ & $\opz \opz \identity$ &
$\opx \opx \opx$ & $\opy \opx \opz$ &
$\opy \opy \opx$ & 1 \\
\hline
7 & $\opy \opy \opy$ & $\opx \opz \opy$ &
$\opx \opy \opx$ & $\opz \opx \identity$ &
$\identity \opx \opz$ & $\opy \opz \opx$ &
$\opz \identity \opz$ & 1 \\
\hline
8 & $\opz \opy \opy$ & $\opx \opz \identity$ &
$\opx \identity \opx$ & $\opy \opx \opy$ &
$\identity \opz \opx$ & $\opz \opx \opz$ &
$\opy \opy \opz$ & 1 \\
\hline
9 & $\opz \opy \identity$ & $\opx \opx \identity $ &
$\identity \identity \opz$ & $\opy \opz \identity$ &
$\opx \opx \opz$ & $\opy \opz \opz$ &
$\opz \opy \opz$ & 2 \\
\hline
\end{tabular}
\caption{A local unitary rotation of Table 1 defining a
``different'', but isomorphic (2,3,4) MUB.}
\label{table2}
\end{table}

Now we apply the \red{nonlocal} unitary operator $\identity
\otimes \opcz$ to the three qubits (where, evidently, the operator
is only nonlocal in the four-dimensional subsystem constituting
the rightmost two qubits). The local transformation we performed
above on the first qubit will of course not change the structure
of the MUB, not even after subsequently applying the operator
$\identity \ot \opcz$. The reason we made this local
transformation is only to facilitate a comparison with \red{the
construction in Ref.~\cite{Lawrence}}. We note that $\opcz
(\identity \ot \opx) \opcz^\dagger = \opcz (\identity \ot \opx)
\opcz = \opz \ot \opx$, and therefore $\opcz (\opz \ot \opx)
\opcz^\dagger = \identity \ot \opx $. The corresponding
transformations for the other product operators are \beq
\begin{array}{ccc}
\opx \ot \identity & \leftrightarrow &
\opx \ot \opz , \\
\identity \ot \opy & \leftrightarrow  &
\opz \ot \opy ,\\
\opy \ot \identity & \leftrightarrow &
\opy \ot \opz .
\\
\end{array}
\eeq
From these, the remaining relations
\beq
\begin{array}{ccc}
\red{\opx \ot \opx \ & \leftrightarrow &
\opy \ot \opy } \\
\opx \ot \opy & \leftrightarrow & \opy  \ot \opx \\
\end{array}
\eeq follow. Hence, applying this transformation to
Table~\ref{table2}, will result in Table~\ref{table3}. From the
unitarity of $\opcz$ it follows that all inner products between
the eigenstates of the simultaneous eigenvectors of the operators
in the same or in different rows of the two tables will be
identical. We can therefore be confident that Table~\ref{table3}
corresponds to a set of MUBs.

\begin{table}[h]
\begin{tabular}{||c||c|c|c|c|c|c|c||c||} \hline
1 & $\opx \identity \identity$ & $\identity \opz \opy$ &
$\identity \opy \opz$ & $\opx \opz \opy$ &
$\identity \opx \opx$ & $\opx \opx \opx$ &
$\opx \opy \opz$ & 2 \\
\hline
2 & $\opy \identity \identity$ & $\identity \opx \opz $ &
$\identity \opz \opx$ & $\opy \opx \opz$ &
$\identity \opy \opy$ & $\opy \opy \opy$ &
$\opy \opz \opx$ & 2 \\
\hline
3 & $\opz \identity \identity$ & $\identity \opy \opx$ &
$\identity \opx \opy$ & $\opz \opy \opx$ &
$\identity \opz \opz$ & $\opz \opz \opz$ &
$\opz \opx \opy$ & 2\\
\hline
4 & $\opy \opz \opy$ & $\identity \opz \identity$ &
$\opx \identity \opz $ & $\opy \identity \opy$ &
$\opx \opz \opz$ & $\opz \identity \opx$ &
$\opz \opz \opx$ & 2 \\
\hline
5 & $\opy \opy \opz$ & $\opx \opy \opx$ &
$\identity \opy \identity$ & $\opz \identity \opy$ &
$\opx \identity \opx$ & $\opz \opy \opy$ &
$\opy \identity \opz$ & 2 \\
\hline
6 & $\opz \opz \opy$ & $\identity \identity \opy$ &
$\opx \opy \identity$ & $\opz \opz \identity$ &
$\opx \opy \opy$ & $\opy \opx \identity$ &
$\opy \opx \opy$ & 2 \\ \hline
7 & $\opy \opx \opx$ & $\opx \identity \opy$ &
$\opx \opx \opy$ & $\opz \opx \opz$ &
$\identity \opx \identity$ & $\opy \identity \opx$ &
$\opz \identity \opz$ & 2 \\
\hline
8 & $\opz \opx \opx$ & $\opx \opz \identity$ &
$\opx \opz \opx$ & $\opy \opy \opx$ &
$\identity \identity \opx$ & $\opz \opx \identity$ &
$\opy \opy \identity$ & 2 \\
\hline
9 & $\opz \opy \opz$ & $\opx \opx \opz $ &
$\identity \identity \opz$ & $\opy \opz \identity$ &
$\opx \opx \identity$ & $\opy \opz \opz$ & $\opz
\opy \identity$ & 2 \\ \hline
\end{tabular}
\caption{Nine sets of operators defining a (0,9,0) MUB.}
\label{table3}
\end{table}

However, this set of MUBs represents a different
entanglement structure, because here every basis
is biseparable. In our nomenclature it is a (0,9,0) set.
In fact it is the same table (with some rows interchanged)
\red{as Fig.~4 in Ref.~\cite{Lawrence}.}

We can now continue and apply the operator $\opcz \ot
\identity$ to Table~\ref{table2}, above. Again the
set of simultaneous eigenstates \red{of} the operators
in any row will define a complete basis, and the
set of bases will form a MUB. The result can be
seen in Table~\ref{table4}.

\begin{table}[h]
\begin{tabular}{||c||c|c|c|c|c|c|c||c||} \hline
1 & $\opx \opz \identity$ & $\identity \identity \opy$ & $\opz \opy
\identity$ & $\opx \opz \opy$ & $\opz \opy \opy$ & $\opy \opx \opy$
& $\opy \opx \identity$ & 2\\ \hline 2 & $\opy \opz \identity$ &
$\opz \opx \identity $ & $\identity \identity \opx$ & $\opx \opy
\identity$ &
$\opz \opx \opx$ & $\opx \opy \opx$ & $\opy \opz \opx$ & 2\\
\hline 3 & $\opz \identity \identity$ & $\opz \opx \opy$ & $\opz
\opy \opx$ & $\identity \opx \opy$ &
$\identity \opz \opz$ & $\opz \opz \opz$ & $\identity \opy \opx$ & 2\\
\hline 4 & $\opy \opz \opy$ & $\identity \opz \identity$ & $\opx
\opz \opz $ & $\opy \identity \opy$ &
$\opx \identity \opz$ & $\opz \opz \opx$ & $\opz \identity \opx$ & 2\\
\hline 5 & $\opx \opx \identity$ & $\opy \opy \opy$ & $\opz \opy
\opz$ & $\opz \opz \opy$ &
$\opx \identity \opx$ & $\identity \opx \opx$ & $\opy \opz \opz$ & 1\\
\hline 6 & $\opz \identity \opy$ & $\identity \opz \opy$ & $\opy
\opx \opz$ & $\opz \opz \identity$ & $\opy \opy \opx$ & $\opx \opy
\opz$ & $\opx \opx \opx$ & 1\\ \hline 7 & $\opx \opx \opy$ & $\opx
\identity \opy$ & $\opy \opx \opx$ & $\identity \opx
\identity$ & $\opz \opx \opz$ & $\opy \identity\opx$ & $\opz \identity \opz$ & 2\\
\hline 8 & $\identity \opy \opy$ & $\opx \identity \identity$ &
$\opx \opz \opx$ & $\opx \opy \opy$ & $\identity \opz
\opx$ & $\identity \opx \opz$ & $\opx \opx \opz$ & 2\\
\hline 9 & $\identity \opy \identity$ & $\opy \opy \identity $ &
$\identity \identity \opz$ & $\opy \identity \identity$ & $\opy \opy
\opz$ & $\opy \identity \opz$ & $\identity \opy \opz$ & 3\\
\hline
\end{tabular}
\caption{Nine sets of operators defining a (1,6,2) MUB.}
\label{table4}
\end{table}
This \red{yields} a $(1,6,2)$ MUB. That is, only one of the bases
\red{is} fully separable. This construction is novel\red{:} it is
neither a structure of the Fields' and Wootter's type, nor is it
one of Lawrence {\it et al}'s two structures. We note from the
tables above, that there are nine operators \red{containing two}
identity operators and \red{twenty seven}  containing a single
identity operator. In each of $s$ sets of operators defining a
fully separable basis (i.e., in each of $s$ rows), there are three
entries with two identity operators each. In each such set there
are also three entries with a single identity operator. Each of
$b$ operator sets of defining a biseparable basis contains one
operator \red{with} two identities, and three operators with a
single identity. Finally, the $n$ sets of operators defining
nonseparable bases contain no operators with two identities, and
three operators with a single identity. \red{In consequence, we
have} the equations \beqa
\red{& 3 s + b = 9 , & \nonumber \\
&& \\
& 3 (s + b + n) = 27 ,} & \nonumber \eeqa for all nonnegative
integers (smaller or equal to nine)\red{, which yield} the four
solutions $\{ (2,3,4), (0,9,0), (1,6,2), (3,0,6) \}$. We conclude
that, so far, we have derived explicit constructions for the first
three structures, and have one more left to construct.

Before doing that, we make a small digression and note that for
three {\em qutrits} \cite{Lawrence 2}, similar considerations lead
to the conclusion that in this 27-dimensional Hilbert space, with 28
MUBs, there exists five MUB structures, namely the set $\{(0,12,16),
(1,9,18), (2,6,20), (3,3,22), (4,0,24) \}$.

Now we return to the three qubit space. The last possible
\red{$(3,0,6)$ structure} can be \red{built up} in the following
way: Take Table~\ref{table4} and perform the transformation $y
\leftrightarrow z$ on the leftmost two qubits and the
transformation $x \leftrightarrow y$ on the rightmost qubit.
\red{As} we are only going to transform the two rightmost qubits
in a nonlocal fashion, the transformation on the leftmost qubit is
only to yield a table that is identical to one of \red{the
constructions in Ref.~\cite{Lawrence}}. The operator
\red{performing} the transformation $x \leftrightarrow y$ is \beq
\opur =
 \left (
\begin{array}{cc}
e^{i \pi/4} &  0  \\
0 & e^{-i \pi/4}
\end{array}
\right ) .
\eeq
The result of the local transformations
$\opuc \ot \opuc \ot \opur$ is \red{shown in }
Table~\ref{table5}.

\begin{table}[h]
\begin{tabular}{||c||c|c|c|c|c|c|c||c||} \hline
1 & $\opx \opy \identity$ & $\identity \identity \opx$ & $\opy \opz
\identity$ & $\opx \opy \opx$ & $\opy \opz \opx$ & $\opz \opx \opx$
& $\opz \opx \identity$ & 2\\ \hline 2 & $\opz \opy \identity$ &
$\opy \opx \identity $ & $\identity \identity \opy$ & $\opx \opz
\identity$ &
$\opy \opx \opy$ & $\opx \opz \opy$ & $\opz \opy \opy$ & 2\\
\hline 3 & $\opy \identity \identity$ & $\opy \opx \opx$ & $\opy
\opz \opy$ & $\identity \opx \opx$ &
$\identity \opy \opz$ & $\opy \opy \opz$ & $\identity \opz \opy$ & 2\\
\hline 4 & $\opz \opy \opx$ & $\identity \opy \identity$ & $\opx
\opy \opz $ & $\opz \identity \opx$ &
$\opx \identity \opz$ & $\opy \opy \opy$ & $\opy \identity \opy$ & 2\\
\hline 5 & $\opx \opx \identity$ & $\opz \opz \opx$ & $\opy \opz
\opz$ & $\opy \opy \opx$ &
$\opx \identity \opy$ & $\identity \opx \opy$ & $\opz \opy \opz$ & 1\\
\hline 6 & $\opy \identity \opx$ & $\identity \opy \opx$ & $\opz
\opx \opz$ & $\opy \opy \identity$ & $\opz \opz \opy$ & $\opx \opz
\opz$ & $\opx \opx \opy$ & 1\\ \hline 7 & $\opx \opx \opx$ & $\opx
\identity \opx$ & $\opz \opx \opy$ & $\identity \opx
\identity$ & $\opy \opx \opz$ & $\opz \identity\opy$ & $\opy \identity \opz$ & 2\\
\hline 8 & $\identity \opz \opx$ & $\opx \identity \identity$ &
$\opx \opy \opy$ & $\opx \opz \opx$ & $\identity \opy
\opy$ & $\identity \opx \opz$ & $\opx \opx \opz$ & 2\\
\hline 9 & $\identity \opz \identity$ & $\opz \opz \identity $ &
$\identity \identity \opz$ & $\opz \identity \identity$ & $\opz \opz
\opz$ & $\opz \identity \opz$ & $\identity \opz \opz$ & 3\\
\hline
\end{tabular}
\caption{A table, isomorphic to Table 4, defining a (1,6,2) MUB.}
\label{table5}
\end{table}

\begin{figure}[h]
\center {\includegraphics[width=85mm]{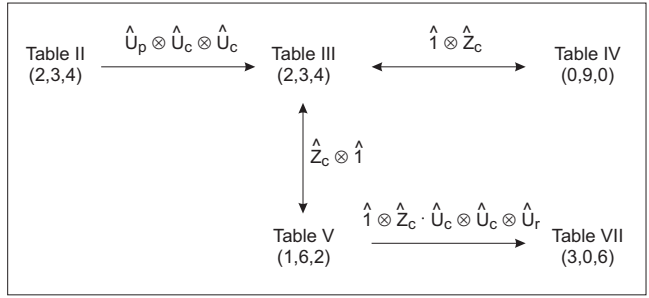}}
\caption{The operational relationship between the different MUB
constructions} \label{Fig:1}
\end{figure}

If we subsequently transform Table~\ref{table5} \red{with}
$\identity \otimes \opcz$, we get the last of the possible
MUB constructions in this eight-dimensional space,
\red{which is reproduced} in Table~\ref{table6}.
\begin{table}[h]
\begin{tabular}{||c||c|c|c|c|c|c|c||c||} \hline
1 & $\opx \opy \opz$ & $\identity \opz \opx$ & $\opy \opz \identity$
& $\opx \opx \opy$ & $\opy \identity \opx$ & $\opz \opy \opy$ &
$\opz \opx \opz$ & 1\\ \hline 2 & $\opz \opy \opz$ & $\opy \opx \opz
$ & $\identity \opz \opy$ & $\opx \opz \identity$ &
$\opy \opy \opx$ & $\opx \identity \opy$ & $\opz \opx \opx$ & 1\\
\hline 3 & $\opy \identity \identity$ & $\opy \opy \opy$ & $\opy
\identity \opy$ & $\identity \opy \opy$ &
$\identity \opy \identity$ & $\opy \opy \identity$ & $\identity \identity \opy$ & 3\\
\hline 4 & $\opz \opx \opy$ & $\identity \opy \opz$ & $\opx \opy
\identity $ & $\opz \opz \opx$ &
$\opx \identity \opz$ & $\opy \opx \opx$ & $\opy \opz \opy$ & 1\\
\hline 5 & $\opx \opx \opz$ & $\opz \identity \opx$ & $\opy \opz
\opz$ & $\opy \opx \opy$ &
$\opx \opz \opy$ & $\identity \opy \opx$ & $\opz \opy \identity$ & 1\\
\hline 6 & $\opy \opz \opx$ & $\identity \opx \opy$ & $\opz \opx
\identity$ & $\opy \opy \opz$ & $\opz \identity \opy$ & $\opx \opz
\opz$ & $\opx \opy \opx$ & 1\\ \hline 7 & $\opx \opy \opy$ & $\opx
\opz \opx$ & $\opz \opy \opx$ & $\identity \opx
\opz$ & $\opy \opx \identity$ & $\opz \opz \opy$ & $\opy \identity \opz$ & 1\\
\hline 8 & $\identity \identity \opx$ & $\opx \identity \identity$ &
$\opx \opx \opx$ & $\opx \identity \opx$ & $\identity \opx
\opx$ & $\identity \opx \identity$ & $\opx \opx \identity$ & 3\\
\hline 9 & $\identity \opz \identity$ & $\opz \opz \identity $ &
$\identity \identity \opz$ & $\opz \identity \identity$ & $\opz \opz
\opz$ & $\opz \identity \opz$ & $\identity \opz \opz$ & 3\\
\hline
\end{tabular}
\caption{Nine sets of operators defining a (3,0,6) MUB.}
\label{table6}
\end{table}
This table is a $(3,0,6)$ MUB. It is, in fact, exactly the
\red{same} (with some rows permuted) as \red{in Fig.~2
of Ref.~\cite{Lawrence}}. The possibilities in the
eight-dimensional space \red{are} now exhausted\red{:}
No \red{other MUBs} with different entanglement
structures can be constructed. We summarize the
operational relationship between the different \red{MUB
structures} one can construct in Fig.~\ref{Fig:1}.

\section{MUB structures for four qubits}

With four qubits, the \red{MUBs} can take one of five different
forms with respect to their separability. We have fully separable
bases, triseparable bases ($2 \times 2 \times 4$), two kinds of
biseparable bases (one that factors $2 \times 8$ and the other
that factors $4 \times 4$), and finally nonseparable bases. If we
follow \red{once more the algorithm given in Ref.~\cite{Klimov}},
but \red{write explicitly} only the first four columns \red{of
each} basis to \red{save} space, we get the following table:

\begin{table}[h]
\begin{tabular}{||c||c|c|c|c||c||} \hline
1 & $\opz \opzii \opziii \opziv$ & $\opz \opzii \identityiii \opziv
$ & $\opz \opzii \opziii \identityiv$ & $\opz \identityii
\identityiii \identityiv$ & 4\\
\hline 2 & $\opx \opxii \opxiii \opxiv$ & $\opx \identityii
\identityiii \identityiv$ & $\identity \opxii \identityiii
\identityiv$ & $\opx \opxii \identityiii \opxiv$ & 4\\
\hline 3 & $\opy \opyii \opyiii \opyiv$ & $\opy \opzii \identityiii
\opziv$ & $\opz \opyii \opziii \identityiv$ & $\opy \opxii
\identityiii \opxiv$ & 1\\
\hline 4 & $\opy \opyii \opxiii \opyiv$ & $\opy \opzii \opziii
\identityiv$ & $\opz \opxii \identityiii \identityiv $ & $\opx
\opyii \opziii \opyiv$ & 2\\
\hline 5 & $\opy \opyii \opyiii \opxiv $ & $\opy \identityii
\identityiii \identityiv $ & $\identity \opyii \opziii \opziv $ &
$\opy \opxii \opziii \opxiv $ & 2\\
\hline 6 & $\opy \opxii \opxiii \opxiv $ & $\opx \opzii \opziii
\opziv $ & $\opz \opxii \opziii \identityiv $ & $\opx \opyii
\identityiii \opxiv $ & 1\\
\hline 7 & $\opx \opyii \opyiii \opyiv $ & $\opy \identityii \opziii
\identityiv $ & $\identity \opyii \identityiii \identityiv $ & $\opy
\opyii \identityiii \opxiv$ & 2\\
\hline 8 & $\opy \opxii \opyiii \opxiv $ & $\opx \opzii \identityiii
\identityiv$ & $\opz \opyii \identityiii \identityiv $ & $\opy
\opxii \opziii \opyiv $ & 2B\\
\hline 9 & $\opx \opyii \opxiii \opxiv $ & $\opy \opzii \identityiii
\identityiv $ & $\opz \opxii \opziii \opziv $ & $\opx \opxii
\identityiii \opyiv $ & 1\\
\hline 10 & $\opy \opyii \opxiii \opxiv $ & $\opy \identityii
\opziii \opziv $ & $\identity \opxii \identityiii \opziv $ & $\opx
\opyii \identityiii \opyiv $ & 1\\
\hline 11 & $\opy \opxii \opyiii \opyiv $ & $\opx \identityii
\identityiii \opziv $ & $\identity \opyii \identityiii \opziv $ &
$\opy \opxii \identityiii \opyiv $ & 2\\
\hline 12 & $\opx \opxii \opxiii \opyiv $ & $\opx \opzii
\identityiii \opziv $ & $\opz \opxii \identityiii \opziv $ & $\opx
\opxii \opziii \opxiv $ & 1\\
\hline 13 & $\opx \opyii \opxiii \opyiv $ & $\opy \identityii
\identityiii \opziv $ & $\identity \opxii \opziii \identityiv $ &
$\opx \opxii \opziii \opyiv $ & 2B\\
\hline 14 & $\opy \opxii \opxiii \opyiv $ & $\opx \identityii
\opziii \identityiv $ & $\identity \opxii \opziii \opziv $ & $\opx
\opyii \opziii \opxiv $ & 1\\
\hline 15 & $\opx \opxii \opyiii \opxiv $ & $\opx \identityii
\opziii \opziv $ & $\identity \opyii \opziii \identityiv $ & $\opy
\opyii \opziii \opyiv $ & 1\\
\hline 16 & $\opx \opxii \opyiii \opyiv $ & $\opx \opzii \opziii
\identityiv $ & $\opz \opyii \opziii \opziv $ & $\opy \opyii
\identityiii \opyiv $ & 1\\
\hline 17 & $\opx \opyii \opyiii \opxiv $ & $\opy \opzii \opziii
\opziv $ & $\opz \opyii \identityiii \opziv $ & $\opy \opyii \opziii
\opxiv $ & 1\\\hline
\end{tabular}
\caption{Seventeen sets of four operators defining a (2,0,4,2,9)
MUB.} \label{table16:1}
\end{table}

The remaining eleven columns of the table can be generated through
the relation $O_{r,c} = O_{r,c-4} O_{r,c-1}$. The 2 in the last
column indicates a basis biseparable in a $2 \times 8$ space,
while 2B indicate a basis biseparable in a $4 \times 4$ space.
This basis will be \red{denoted as} a (2,0,4,2,9) MUB, referring
how many of the bases that are fully separable, triseparable,
biseparable (in a $2 \times 8$ and in a $4 \times 4$ space,
respectively), and nonseparable. We now apply the operator \beq
\identity \ot \identity \ot \opcz \red{\cdot} \identity \ot
\identity \ot \opuc \ot \opuc, \eeq \red{which} first locally
rotates the two rightmost qubits so that $\opy \leftrightarrow
\opz$ and then entangles (or disentangles) the same two qubits.
The result is Table~\ref{table16:2}.

\begin{table}[h]
\begin{tabular}{||c||c|c|c|c||c||} \hline
1 & $\opz \opzii \opxiii \opxiv$ & $\opz \opzii \opziii \opyiv $ &
$\opz \opzii \opyiii \opziv$ & $\opz \identityii \identityiii
\identityiv$ & 3\\
\hline 2 & $\opx \opxii \opyiii \opyiv$ & $\opx \identityii
\identityiii \identityiv$ & $\identity \opxii \identityiii
\identityiv$ & $\opx \opxii \opziii \opxiv $ & 3\\
\hline 3 & $\opy \opyii \opziii \opziv$ & $\opy \opzii \opziii
\opyiv$ & $\opz \opyii \opyiii \opziv$ & $\opy \opxii \opziii \opxiv
$ & 2B\\
\hline 4 & $\opy \opyii \opxiii \identityiv$ & $\opy \opzii \opyiii
\opziv$ & $\opz \opxii \identityiii \identityiv $ & $\opx \opyii
\opyiii \identityiv$ & 2\\
\hline 5 & $\opy \opyii \identityiii \opxiv $ & $\opy \identityii
\identityiii \identityiv $ & $\identity \opyii \opxiii \opxiv $ &
$\opy \opxii \opxiii \opyiv $ & 3\\
\hline 6 & $\opy \opxii \opyiii \opyiv $ & $\opx \opzii \opxiii
\opxiv $ & $\opz \opxii \opyiii \opziv $ & $\opx \opyii \opziii
\opxiv $ & 2B\\
\hline 7 & $\opx \opyii \opziii \opziv $ & $\opy \identityii \opyiii
\opziv $ & $\identity \opyii \identityiii \identityiv $ & $\opy
\opyii \opziii \opxiv $ & 2\\
\hline 8 & $\opy \opxii \identityiii \opxiv $ & $\opx \opzii
\identityiii \identityiv$ & $\opz \opyii \identityiii \identityiv $
& $\opy \opxii \opyiii \identityiv $ & 3\\
\hline 9 & $\opx \opyii \opyiii \opyiv $ & $\opy \opzii \identityiii
\identityiv $ & $\opz \opxii \opxiii \opxiv $ & $\opx \opxii
\identityiii \opziv $ & 1\\
\hline 10 & $\opy \opyii \opyiii \opyiv $ & $\opy \identityii
\opxiii \opxiv $ & $\identity \opxii \opziii \opyiv $ & $\opx \opyii
\identityiii \opziv $ & 1\\
\hline 11 & $\opy \opxii \opziii \opziv $ & $\opx \identityii
\opziii \opyiv $ & $\identity \opyii \opziii \opyiv $ & $\opy \opxii
\identityiii \opziv $ & 2\\
\hline 12 & $\opx \opxii \opxiii \identityiv $ & $\opx \opzii
\opziii \opyiv $ & $\opz \opxii \opziii \opyiv $ & $\opx \opxii
\opxiii \opyiv $ & 2\\
\hline 13 & $\opx \opyii \opxiii \identityiv $ & $\opy \identityii
\opziii \opyiv $ & $\identity \opxii \opyiii \opziv $ & $\opx \opxii
\opyiii \identityiv $ & 1\\
\hline 14 & $\opy \opxii \opxiii \identityiv $ & $\opx \identityii
\opyiii \opziv $ & $\identity \opxii \opxiii \opxiv $ & $\opx \opyii
\opxiii \opyiv $ & 1\\
\hline 15 & $\opx \opxii \identityiii \opxiv $ & $\opx \identityii
\opxiii \opxiv $ & $\identity \opyii \opyiii \opziv $ & $\opy \opyii
\opyiii \identityiv $ & 1\\
\hline 16 & $\opx \opxii \opziii \opziv $ & $\opx \opzii \opyiii
\opziv $ & $\opz \opyii \opxiii \opxiv $ & $\opy \opyii \identityiii
\opziv $ & 1\\
\hline 17 & $\opx \opyii \identityiii \opxiv $ & $\opy \opzii
\opxiii \opxiv $ & $\opz \opyii \opziii \opyiv $ & $\opy \opyii
\opxiii \opyiv $ & 1\\\hline
\end{tabular}
\caption{Seventeen sets of four operators defining a (0,4,4,2,7)
MUB.} \label{table16:2}
\end{table}

Before venturing further, it is instructive to see how
many different MUB structures there are in the four-qubit
space. Again we use the fact \red{that} identity operators
play a special role in defining the separability of the
bases. \red{Table~\ref{Table:identities}} shows how many
products of three, two\red{,} and single identity
operators define a basis of a certain kind.
The ordering of the multiple identities plays no
role.

\begin{table}[h]
\begin{tabular}{||c||c|c|c|} \hline
Basis separability& $\identity \ot \identity \ot \identity$ &
$\identity \ot
 \identity$ & $\identity$ \\ \hline
4 & 4  & 6 & 4 \\ \hline 3 & 2 & 4 & 6\\ \hline 2 & 1 & 3 & 7\\
\hline 2B & 0 & 6 & 0\\ \hline 1 & 0 & 2 & 8\\ \hline \hline
Available entries& 12 & 54 & 108 \\ \hline
\end{tabular}
\caption{The separability of the bases (left column) and the number
of triplets, pairs, and single identity operators contained in the
basis defining operator set.} \label{Table:identities}
\end{table}

Solving the three equations for the different number of identity
operators, we find sixteen different MUB structures: (3,0,0,2,12),
(2,0,4,2,9), (2,1,2,2,10), (2,2,0,2,11), (1,0,8,2,6), (1,1,6,2,7),
(1,2,4,2,8), (1,3,2,2,9), (1,4,0,2,10), (0,0,12,2,3),
(0,1,10,2,4), (0,2,8,2,5), (0,3,6,2,6), (0,4,4,2,7), (0,5,2,2,8),
and (0,6,0,2,9). Of these\red{,} we have explicitly given the
\red{tables} for the $(2,0,4,2,9)$ and the $(0,4,4,2,7)$
structures. Deriving the transformations between any two of the
sixteen structures goes beyond the scope of this paper. However,
\red{note that} applying the operator $\identity \ot \opcz \ot
\identity$ to the entries of Table~\ref{table16:2}, will yield a
table with the MUB structure (0,3,6,2,6). Moreover, \red{using}
instead the operator \beq \opcz \ot \identity \ot \identity \cdot
\opuc \ot \opuc \ot \identity \ot \identity \eeq \red{in} each
entry of Table~\ref{table16:2} will yield a (1,2,4,2,8) MUB
structure. Since we know that a sequence of C-NOT or
controlled-$Z$ operations, together with local unitary rotations,
suffice to make any entanglement transformation on qubits, it is
clear that similar transformations will yield the whole set of
different MUB structures, starting from the
\red{Table~\ref{table16:1}.}

\section{Conclusions}

\red{For one} and two qubits, all possible MUBs have the same
structure with the respect of entanglement. For \red{more}
qubits\red{,} the situation \red{is more involved}: four different
MUB structures \red{appear for three qubits and seventeen MUB
structures for four qubits}. The difference between \red{these}
structures lies in how the bases are entangled. For both three and
four qubits, MUBS exist that have 3, 2, 1 and no fully separable
basis set(s). For three qubits it is possible to find one MUB that
have no fully \red{nonseparable} bases. This is no longer possible
in the four qubit case.

In a quantum protocol relying on MUBs, the entanglement
structure of the MUB is usually inconsequential. What
counts is usually only the mutual unbiasedness,
not the separability  of the bases. Experimentally,
however, it may be easier to generate one set
of bases rather than another. Some of the bases can
be generated locally, \red{accessing} each qubit
separately. However, as we have shown, when
several qubits are involved, most bases are entangled
in one way or another, requiring joint operations on
the qubits. In \red{this} paper we have tried to
delineate the possible MUB structures for up to
four qubits. The method we have employed can of course
be extended to any number of qubits, \red{although} the
complexity and variety of bases grows very rapidly
with the number of qubits.

\acknowledgments
This work was supported by the Swedish Foundation
for International Cooperation in Research and Higher Education
(STINT), the Swedish Foundation for Strategic Research (SSF), and
the Swedish Research Council (VR).

\end{document}